\documentclass{PoS}

\title{ On
Confinement, Chiral Symmetry Breaking, and the U$_A$(1) anomaly in
Functional Approaches}

\ShortTitle{Confinement \& $\chi$SB in Fun.~Approaches}

\author{\speaker{Reinhard Alkofer}\\
Institut f\"ur Physik,
Karl-Franzens-Universit\"at,
Universit\"atsplatz 5,
A-8010 Graz, Austria\\
E-mail: \email{reinhard.alkofer@uni-graz.at}}

\abstract{The so-called decoupling and scaling solutions of functional equations
of Landau gauge Yang-Mills theory are briefly reviewed. In both types of
solutions the positivity violation seen in the gluon propagator is taken as an
indication of gluon confinement. In the scaling solution the
resulting infrared singularities of the quark-gluon vertex 
are responsible for the linear potential between static quarks 
and are therefore signaling quark confinement. 
A corresponding description of the U$_A$(1) anomaly
in functional approaches is only known for the scaling solution.
Nevertheless, it seems puzzling at first sight that quark confinement is related
to the dynamical and anomalous breaking of chiral symmetry in a self-consistent
manner: One  obtains either all these phenomena or none. For the scaling
solution also fundamental scalar fields are confined. This provides evidence
that within functional approaches static confinement is an universal property of
the gauge sector even though it is formally represented in the functional
equations of the matter sector. }

\FullConference{The many faces of QCD\\
		November 2-5, 2010\\
		Gent Belgium}

\begin{document}

\section{Introduction: A note on the axial anomaly}

This workshop is called {\it The many faces of QCD}, and actually most
participants really experienced their work with QCD being multifaceted. First of
all, most successes of QCD are related to processes with high-momentum transfer
in which asymptotic freedom \cite{Gross:1973id,Politzer:1973fx} enables the use
of perturbation theory. On the other hand, although QCD was invented 37 years
ago \cite{Fritzsch:1973pi} we only start to
understand  its infrared regime where we face all kind of strong-interaction
phenomena, most prominently confinement, anomalous and dynamical chiral symmetry
breaking, and the formation of relativistic bound states. 

With respect to the anomaly I want to recall a seminal result \cite{Vafa:1983tf}
which may be loosely summarized as follows: the axial U(1) symmetry is always 
anomalously broken in vector-like gauge theories with  vacuum angle $\Theta =
0$. One possibility to explain this anomaly rests on the existence of 
quark zero modes in topologically non-trivial fields 
\cite{'tHooft:1976fv,Leutwyler:1992yt}: A  random distribution of (not
necessarily integer) winding number spots leads to a non-vanishing topological
susceptibility in the thermodynamic limit. Via the index theorem one can then
associate percolating quark zero modes, and they eventually cause the anomalous
breaking of the axial U(1) symmetry.

As this explanation is so overwhelmingly successful the question arises whether
it is the only existing one. And if another one is available, do these several
explanations exclude each other? Here an historic example might be helpful:
Everybody of us remembers from his graduate lectures how to derive Bloch waves
in a periodic potential by employing the Schr\"odinger  equation. A  short look
in Sidney Coleman's Erice Lectures ``The uses of instantons''
\cite{Coleman:1978ae}, however, tells us how to achieve the same by instanton
calculus techniques. Of course, nobody of us would ever dare to believe that
one would have to add instantons to the Schr\"odinger  equation to obtain Bloch
waves. The Schr\"odinger  equation and instanton calculus are simply two
different techniques to obtain the same physical result. On the other hand, one
can hear quite often the opinion that one has to add to some non-perturbative
techniques (as {\it e.g.\/} functional equations) the instantons (or other
topologically non-trivial field configurations)  by hand to obtain a
non-vanishing topological susceptibility. The above comparison should, however,
elucidate that adding to a consistent approach some other ingredients results
in an incorrect treatment.

Accepting this, the following question arises: Where is the topological
susceptibility encoded in an approach based on %non-perturbative 
Green's
functions? The decisive hint originates from the seventies 
\cite{Kogut:1973ab} (see also \cite{vonSmekal:1997dq}).  Rephrasing this old
result in modern language one may state that momentum-space Green's function
can reflect the topological susceptibility only in their infrared behaviour
because only these are related to the boundary conditions in (Euclidean)
space-time. 

Emphasizing with this introductory remark the special role of the axial anomaly
in our understanding of QCD let me give a short outline of the following
sections: After shortly reviewing the knowledge on the infrared structure of
Landau gauge Yang-Mills theory I will focus on the positivity violation of the
gluon propagator and potential implications for its analytic structure. For
fundamental charges the corresponding gluon-matter vertex functions are
analysed. Hereby it is demonstrated that the quark-gluon vertex may play a key
role in the issue of quark confinement. The quark-gluon vertex is hereby
twofold related in self-consistent manner to dynamical chiral symmetry breaking
(D$\chi$SB): On the one hand, its strength triggers D$\chi$SB, on the other hand
it is subject of D$\chi$SB and contains components which are only possible due
to D$\chi$SB. A study of the infrared properties of fundamentally charged scalars
provides evidence that within functional approaches static confinement is an
universal property of the gauge sector even though it is formally represented in
the functional equations of the matter sector.\footnote{However, one has to note
that the corresponding lattice results reported by Axel Maas in this workshop
\cite{Maas:2011yx} do not corroborate this evidence.} Last but not least, I will
return to the question of the description of the axial anomaly within functional
approaches.

\section{Infrared Structure of Landau gauge Yang-Mills theory}

The indefinite metric state space of a Yang-Mills theory  can be classified
according to the properties of the states under BRST transformation, see {\it
e.g.\/} \cite{Kugo:1979gm,Nakanishi:1990qm,vonSmekal:2000pz}. The BRST
cohomology contains the physical states, the unphysical states form quartets.
Such quartets do exist either as perturbative or non-perturbative ones
\cite{Nakanishi:1990qm,Alkofer:2011bb,Alkofer:2011pe}. One important ingredient
in the construction of a BRST quartet generated by transverse gluons is the
fact that a ``mass'' gap in transverse gluon correlations needs to be
generated,  {\it i.e.}, the massless transverse gluon states of perturbation
theory have to dissappear even though they should belong to quartets due to
color antiscreening and superconvergence in QCD
\cite{Oehme:1980ai,Alkofer:2000wg}. Within this formulation one can provide a
clear distinction between the confinement and the Higgs phase: In the former
the colour charge is well-defined in the whole state space, in the latter it
is not. A condition which leads to such a well-defined charge can be shown in 
Landau gauge by standard arguments employing functional equations and
Slavnov-Taylor identities to be equivalent to an infrared enhanced ghost
propagator \cite{Kugo:1995km,Alkofer:2000wg} which in turn then implies an
infrared vanishing gluon propagator 
\cite{vonSmekal:1997is,Watson:2001yv,Zwanziger:2001kw,Lerche:2002ep,Fischer:2002hna,Pawlowski:2003hq,Fischer:2008uz}.\footnote{This
so-called scaling solution of functional equations has been debated quite
intensively recently. One should note, however, that the violation of positivity
for gluons is generally accepted, see {\it e.g.} \cite{Bowman:2007du}.}

The implications of a broken colour charge are quite
straightforward \cite{Nakanishi:1990qm}: In each channel in which the gauge
potential contains an asymptotic massive vector field the global gauge symmetry
generated by the colour charges is spontaneously broken. While this  massive
vector state results to be a BRST-singlet,  the massless Goldstone boson states,
which usually occur in some components of the Higgs field, replace the third
component of the vector field in the elementary quartet and are thus
unphysical. Since the broken charges are BRST-exact, this {hidden}
symmetry breaking is not observable in the Hilbert space
of physical states.
Thus, if the gauge boson is massive it possesses three
degenerate polarization states. Everything else would have been a surprise
because with respect to the representations of the Poincar{\'e} group there are
only two choices:
\begin{itemize}
\item massive and three polarization states, or\\[-8mm]
\item massless and two polarization states.
\end{itemize}

With this remarks in mind let us now analyse the situation in QCD ({\it i.e.},
in the confinement phase) and assume hereby either of the two types of
solutions found in functional equations, namely the scaling one with an
infrared vanishing gluon propagator or the decoupling ones with an infrared
finite gluon propagator, for a description of the latter solutions
see refs.~\cite{Aguilar:2008xm,Boucaud:2008ky} and
references therein. 
\begin{itemize}
\item
An infrared vanishing gluon propagator has a vanishing screening length, the
corresponding screening ``mass'' is thus infinite.  Nevertheless one would
not attribute an infinite gluon mass.
\item
An infrared finite gluon propagator has a finite screening length, the
corresponding screening ``mass'' is therefore finite. However, accepting the
positivity violation of transverse gluons as a fact the question arises in
which sense this relates to a mass.\footnote{The clearest definition of mass in
context of a relativistic quantum field theory is that mass is the square root
of the first quadratic Casimir invariant of the  Poincar{\'e} group,
$m:=\sqrt{P_\mu P^\mu}$.}
\item
Longitudinal and transverse gluons do not belong to the same BRST
representation as there is no doubt that in the confinement phase the
longitudinal gluon belongs to the perturbative elementary quartet. This implies
that the longitudinal gluon stays a massless (unphysical) state. Putting now the
tranverse gluons into the same BRST representation\footnote{BRST multiplets are
degenerate as the BRST charge commutes with the Hamiltonian.}  
clearly contradicts the necessity of generating a ``mass'' gap for the
transverse gluons. 
\item
This is corroborated by the fact that glueballs (which are would-be physical
states in pure Yang-Mills theory) do not contain any contribution of 
longitudinal gluons \cite{Mathieu:2009cc}.
\end{itemize}
The only possible conclusion from this is that longitudinal and transverse
gluons are not in the same representation of the  Poincar{\'e} group. 
A Poincar{\'e} representation for a vector with two polarization states 
is certainly not the representation for a massive vector. 

To summarize this argument: Without the longitudinal polarization as part of
the Poincar{\'e}  representation of the transverse gluons my choice is to
refrain from statements like ``The gluon is massive.'' or phrases like ``the
gluon mass'', and this  independent of what the value of the auto-correlation
function of excitations of transverse gluons at vanishing virtuality $k^2=0$ is.
To my opinion, calling a gluon
``massive'' is confusing the issue of gluon confinement.

\subsection{Infrared Exponents for Gluons and Ghosts}

As already stated the infrared behaviour of the one-parameter family of
decoupling solutions is such that one obtains an infrared finite gluon
propagator and otherwise infrared trivial Green's functions
\cite{Aguilar:2008xm,Boucaud:2008ky,Fischer:2008uz}. The end-point of these
solutions is the scaling solution. The infrared behaviour of all one-particle
irreducible Green's functions in the scaling solution is easily described in the
simplified case with only one external scale $p^2\to 0$: For a function with $n$
external ghost and antighost as well as $m$ gluon legs one obtains
\cite{Alkofer:2004it,Huber:2007kc}
\begin{equation}
\Gamma^{n,m}(p^2) \sim (p^2)^{({n-m})\kappa} .
\end{equation}
This solution fulfills all functional equations  and all
Slavnov-Taylor identities. It verifies the hypothesis of infrared
ghost dominance \cite{Zwanziger:2003cf} and leads to infrared diverging 3- and
4-gluon vertices.

There is only one unique scaling solution with power laws for the Green's
functions \cite{Fischer:2006vf,Fischer:2009tn}. A detailed comparison of both
type of solutions can be found {\it e.g.} in \cite{Fischer:2008uz}, an infrared
analysis for both type of solutions is described {\it e.g.} in Ref.\
\cite{Alkofer:2008jy}. Although almost all lattice calculations of
the gluon propagator favor the decoupling solution it is certainly worthwhile to
study the scaling solution as a theoretical tool. And there is the possibility
that the difference between these solutions depends on nothing else than a
choice of gauge  \cite{Maas:2009se}. The latter interpretation is corroborated
by the fact that lattice studies at strong coupling
\cite{Sternbeck:2008mv,Maas:2009ph,Cucchieri:2009zt} reveal the existence of a
regime where the scaling relation between the gluon and the ghost propagator is
fulfilled, and the corresponding infrared exponent $\kappa$ is very close to the
value determined in a full class of
truncated continuum studies with $\kappa=  \frac{93 - \sqrt{1201}}{98}
 \simeq 0.59535$.

\subsection{Positivity violation of the gluon propagator}

The  positivity violation of the (space-time) propagator of
transverse gluons as predicted by the Oehme--Zimmermann superconvergence
relation \cite{Oehme:1980ai} and
corresponding to the Kugo--Ojima \cite{Kugo:1979gm}
and Gribov--Zwanziger \cite{Zwanziger:2003cf} scenarios
has been a long-standing conjecture for which there is now
compelling evidence, see  {\it e.g.\/} Refs.\ 
\cite{Bowman:2007du,Alkofer:2003jj} and
references therein. The basic features underlying these gluon properties,
are the infrared suppression of correlations of transverse gluons and the
infrared enhancement of ghost correlations as discussed above.
A simple argument given by Zwanziger makes this at least for the scaling
solution obvious: An infrared 
vanishing gluon propagator implies for the space-time gluon
propagator being the Fourier transform of the momentum space gluon propagator:
\begin{equation}
0=D_{gluon}(k^2=0) = \int d^4x \; \;\; \;  D_{gluon}(x) \; .
\end{equation}
\begin{figure}[htbp]
\begin{center}
\includegraphics[width=70mm]{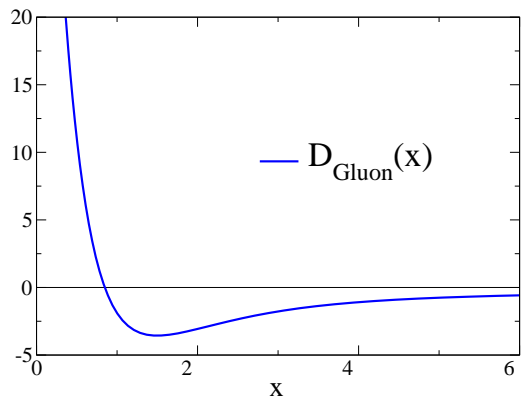}
\label{Gluon-pos}
\caption{The Fourier transform of the scaling solution
 for the gluon propagator.}
\end{center}
\end{figure}
Therefore $D_{gluon}(x)$ has to be negative for some values of $x$. 
Exactly this behaviour is seen in Fig.~1 %\ref{Gluon-pos}  
in which the
Fourier transform of the scaling solution  for the gluon propagator is
displayed.

In order to investigate the analytic structure of the gluon propagator
we first parameterize the running coupling such that the numerical results
for Euclidean scales are  reproduced \cite{Fischer:2003rp}:
\begin{eqnarray}
\alpha_{\rm fit}(p^2) = \frac{\alpha_S(0)}{1+p^2/\Lambda^2_{\tt QCD}}
+ \frac{4 \pi}{\beta_0} \frac{p^2}{\Lambda^2_{\tt QCD}+p^2}
\left(\frac{1}{\ln(p^2/\Lambda^2_{\tt QCD})}
- \frac{1}{p^2/\Lambda_{\tt QCD}^2 -1}\right) 
\end{eqnarray}
with $\beta_0=(11N_c-2N_f)/3$. In this expression the Landau pole has been
subtracted,
it is analytic in the complex $p^2$ plane except the real timelike axis
where the logarithm produces a cut for real $p^2<0$, and it obeys Cutkosky's rule.

The infrared exponent $\kappa$ is an irrational number, and thus  the
gluon propagator possesses a cut on the
negative real $p^2$ axis. It is possible to fit the solution for the gluon
propagator quite accurately without introducing further singularities
in the complex $p^2$ plane.
The fit to the gluon renormalization function \cite{Alkofer:2003jj}
\begin{equation}
Z_{\rm fit}(p^2) = w \left(\frac{p^2}{\Lambda^2_{\tt QCD}+p^2}\right)^{2 \kappa}
 \left( \alpha_{\rm fit}(p^2) \right)^{-\gamma}
 \label{fitII}
\end{equation}
works quite precisely. Hereby $w$ is a normalization parameter, and
$\gamma = (-13 N_c + 4 N_f)/(22 N_c - 4 N_f)$
is the one-loop value for
the anomalous dimension of the gluon propagator.
The discontinuity of (\ref{fitII}) along the cut
vanishes for $p^2\to 0^-$, diverges to $+\infty$ at $p^2=-\Lambda_{\tt QCD}^2$
and goes to zero for $p^2\to \infty$.
The function (\ref{fitII}) contains only four parameters:  the overall magnitude
which due to renormalization properties is arbitrary (it
is determined via the choice of the renormalization scale),   the
scale $\Lambda_{\tt QCD}$,   the infrared exponent $\kappa$ and  the
anomalous dimension of the gluon $\gamma$. The latter two are not
free parameters: $\kappa$ is determined from the infrared properties of the
DSEs and for $\gamma$ its one-loop value is used. Thus we have found a
parameterization of the gluon propagator which has effectively only one
parameter, the scale $\Lambda_{\tt QCD}$.
It is important to note that the gluon propagator possesses a form such
that {\em Wick rotation is possible!}

\section{Quarks/Matter: Confinement vs.~D$\chi$SB \& U$_A$(1) anomaly }

Due to the infrared suppression of the gluon propagator, present in the scaling
{\bf and} in the decoupling solutions, quark confinement (or, generally,
confinement of fundamental charges) cannot be generated by any type of gluon
exchange together with infrared-bounded vertex functions.  Therefore it is
mandatory to study the functional  equations for the quark propagator together
with the one for the quark-gluon vertex in a self-consistent way
\cite{Alkofer:2006gz,Alkofer:2008tt}. An important difference of the quarks as
compared to Yang-Mills fields arises: As the former possess a mass, and as
D$\chi$SB does occur, the quark propagator will always approach a constant in
the infrared.

\subsection{Dynamically induced scalar quark confinement}

The fully dressed quark-gluon vertex consists of twelve linearly
independent Dirac tensors. Half of the coefficient functions would vanish if
chiral symmetry were realized in the Wigner-Weyl mode. From a solution of the
Dyson-Schwinger equations we infer that  these {``scalar''} structures are,
in the chiral limit, generated non-perturbatively together with the dynamical
quark mass function in a self-consistent fashion. This implies the important
result that dynamical chiral symmetry breaking manifests itself not only 
in the propagator but also in the quark-gluon vertex.

From an infrared analysis one obtains an infrared divergent solution for the
quark-gluon vertex such that  Dirac vector and {``scalar''} components of
this  vertex are infrared divergent with exponent $-\kappa - \frac 1 2$ if
either all momenta or the gluon momentum vanish
\cite{Alkofer:2008tt}. A numerical solution of a truncated set of
Dyson-Schwinger equations confirms this infrared behaviour.
The essential components to obtain this solution are the ``scalar'' Dirac
amplitudes of the quark-gluon vertex and the scalar part of the quark
propagator. Both are only present when chiral symmetry is broken, either
explicitely or dynamically.

In order to determine how this self-consistent quark propagator and quark-gluon
vertex solution relates to quark confinement, the anomalous
infrared exponent of the four-quark function is calculated.
The static quark potential can be obtained
from this four-quark one-particle irreducible Green function. In the scaling
solution it
behaves like $(p^2)^{-2}$ for $p^2\to0$ due to the infrared enhancement
of the quark-gluon vertex for vanishing gluon momentum.
Using a well-known relation one obtains for the static quark-antiquark potential
$V({\bf r})$: 
\begin{equation}
V({\bf r}) \ \ \sim \ \ 
 \int \frac{d^3p}{(2\pi)^3} \left. \frac{1}{p^4} \right|_{p^0=0}
  e^{i {\bf p r}}
\ \ \sim \ \ |{\bf r} |
\end{equation}
Therefore the infrared divergence of the quark-gluon vertex, as found in the
scaling solution of the coupled system of Dyson-Schwinger equations, the vertex
overcompensates the infrared suppression of the gluon propagator such that one
obtains a linearly rising potential.

\subsection{Fundamentally charged scalar field}

Given the complications with the many tensor structures for quarks, and given
the cost for fermions on the lattice, it seems natural to use fundamentally
charged scalars as a laboratory to study confinement. In this context the scalar
propagator and the scalar-gluon vertex were investigated on the lattice
\cite{Maas:2010nc,Maas:2011yx} and analytically
\cite{Fister:2010yw,Fister:2010ah,Alkofer:2010tq}.  Different than the quark
Green's functions the tensor structure of the scalar ones is strongly
simplified. Compared to two components in the fermionic propagator, the scalar
propagator features only a single structure. Similarly the vertex depending on
two independent momenta can be decomposed into two tensors (instead of twelve).

A scalar possesses self-interactions and therefore the
number of terms in the Dyson-Schwinger and Functional  Renormalization Group
equations is significantly increased.  For the derivation of the
Dyson-Schwinger equations one may employ the MATHEMATICA package DoDSE
\cite{Alkofer:2008nt}. (A package for Functional Renormalization
Group equations will be published soon \cite{Huber:2011}.)
In the uniform scaling limit, applying the 
constraints on the infrared exponents arising from the comparison of the
inequivalent towers of Dyson-Schwinger and  Functional Renormalization Group
equations \cite{Fischer:2009tn}, the system of equations for the
anomalous exponents simplifies. One obtains the scaling and the decoupling
solutions with an unaltered Yang-Mills sector. In the case of the scaling
solution for a massive scalar, the scalar-gluon vertex can show two distinct
behaviours \cite{Fister:2010yw,Fister:2010ah}.
In the one be discussed further it exhibits the
same infrared exponent as the quark-gluon vertex.

The uniform scaling uncovers only a small part of the potential infrared
enhancements. Vertex functions may also  become divergent when only a subset of
the external momenta vanish. Such kinematic divergences provide a mechanism for
the long-range interaction of quarks as described in the section above. 
It is gratifying to realize that the kinematic divergences of the
scalar-gluon vertex are identical to those of the quark-gluon vertex. These
singularities induce a confining interaction in the four-scalar vertex function
as they did in the case of the four-quark vertex function in the case of scalar
QCD.  Their Fourier transform  leads to a linearly rising static potential.

This result provides the possibility that within functional approaches static
confinement is an universal property of the gauge sector even though it is
formally represented in the functional equations of the matter sector.
Unfortunately, these results are not corroborated by the lattice results, 
see Refs.~\cite{Maas:2010nc,Maas:2011yx} for more details.

\subsection{ U$_A$(1):
$\eta^\prime$ mass from infrared divergent Green functions}

\begin{figure}
\centerline{
  \includegraphics[width=.6\textwidth]{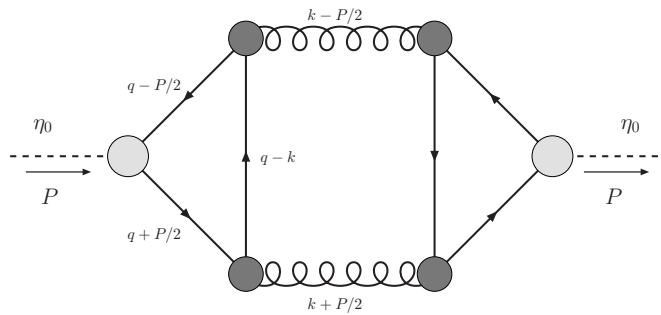}}
  \caption{Contribution to the  $\eta^\prime$ mass due to the infrared
  divergence of the product of quark-gluon vertices and gluon propagator,
   $\Gamma_\mu D^{\mu\nu}
  \Gamma_\nu \propto 1/(k\pm P/2)^4$.}
  \label{diamond}
\end{figure}
Based on purely dimensional arguments \cite{Kogut:1973ab} one can conclude that
the diamond diagram depicted in Fig.\ \ref{diamond} supplies  a non-vanishing
contribution to the mass of the pseudo-scalar flavor-singlet meson in the
chiral limit if the effective one-gluon-exchange diverges with the
gluon-momentum like $1/k^4$. As discussed above this is exactly the behaviour
found in the scaling solution for the  product of two quark-gluon vertices 
and the gluon propagator when the exchanged gluons become soft,
$\Gamma_\mu (p,q;k) D^{\mu\nu} (k)  \Gamma_\nu (r,s;k) \propto 1/k^4$.   
An explicit calculation  \cite{Alkofer:2008et}  verifies the corresponding
generation of a flavor-singlet mass. However, it has to be noted that the 
exchange of more than two gluons also generate contributions to the
$\eta^\prime$ mass. As a matter of fact infinitely many diagrams contribute.
Under this aspect it reassuring that the diagram of  Fig.\ \ref{diamond}
provides the leading contribution. 
Expressing the result in terms of the topological susceptibility $\chi^2$
one obtains $\chi^2 = (160$MeV$)^4$ as compared to the phenomenological value
 $\chi^2 = (180$MeV$)^4$ \cite{Alkofer:2008et}.

In this picture the infrared divergence of the quark-gluon vertex plays an
important role in a confinement-based explanation of the  $\eta^\prime$ mass,
the topological susceptibility and the $U_A(1)$ anomaly. This
provides evidence that the confining field configurations of QCD are 
topologically non-trivial. For example, when removing  center
vortices from a lattice ensemble, the string tension vanishes and
the Landau gauge ghost propagator becomes infrared suppressed
\cite{Gattnar:2004bf}.\footnote{For an introduction to confining field
configurations see {\it e.g.} Refs.\ \cite{Alkofer:2006fu,Greensite:2011zz}.}

The appearance of the correct
infrared singularity in a single Feynman diagram is the case only for the
scaling solution. One may therefore speculate that for a decoupling
solution only a resummation of infinitely many diagrams will be able to
describe the axial anomaly.

\section{Summary}

The unique scaling and the family of decoupling solutions of the functional
equations of Yang-Mills theory have been presented. It has been conjectured that
the appearance of several solutions is related to the choice of gauge 
\cite{Maas:2009se}. This
would especially imply that all these solutions give the same values for
physical observables.

A relatively simple form for the analytic structure of the gluon propagator
has been suggested \cite{Alkofer:2003jj}. It has the remarkable property that it allows a Wick
rotation. 

The quark-gluon vertex plays a double role in dynamical chiral symmetry 
breaking: This vertex triggers and is subject  of the symmetry breaking
\cite{Alkofer:2008tt}.
This results in a quite complicated Dirac structure of the static linearly
rising quark potential. Analytical results for the scaling solution for a
fundamentally charged scalar exist \cite{Fister:2010yw}. 
However, lattice results do not corroborate
them \cite{Maas:2010nc}.

The infrared singularities of the quark-gluon vertex for soft gluon momenta
generate in the scaling scenario an $\eta^\prime$ mass and the 
axial anomaly \cite{Alkofer:2008et}. To my best knowledge, it is yet
unknown how the axial
anomaly is encoded in the elementary Green's functions of the decoupling
solution.

\section*{Acknowledgments}
It is a pleasure to thank the organizers of this highly interesting workshop for
their respective efforts. My special thanks goes to David Dudal for enabling
financial support.
I am grateful to  
%Daniele Binosi, David Dudal, Christian Fischer, Holger Gies, Markus Huber,
%Felipe Llanes-Estrada, Axel Maas, Michael M\"uller-Preu\ss ker, Joannis
%Papavassiliou, Jan Martin Pawlowski, Kai Schwenzer, Silvio Sorella, Andre
%Sternbeck, Lorenz von Smekal, and Daniel Zwanziger
%for helpful discussions and/or 
Christian Fischer, Leo Fister, Markus Huber, Felipe Llanes-Estrada,
Axel Maas,  
Kai Schwenzer and   Lorenz von Smekal for 
collaborations on the research
reported here.

This work was supported in part by the Austrian Science Fund FWF under  Project
No.\ P20592-N16 and by the European Union (HadronPhysics2 project ``Study of
strongly-interacting matter'').

\end{document}